\documentclass[twocolumn,showpacs,preprintnumbers,superscriptaddress,
               amsmath,amssymb,prd]{revtex4}
\usepackage[dvips]{graphicx}

\newcommand{\tr}{\text{tr}}
\newcommand{\Sg}{S_{\text{g}}}
\newcommand{\Sf}{S_{\text{f}}}
\newcommand{\Smf}{S_{\text{mf}}}
\newcommand{\Smag}{S_{\text{f}}^{\text{mag}}}
\newcommand{\fmf}{f_{\text{mf}}}
\newcommand{\Nf}{N_{\text{f}}}
\newcommand{\Nc}{N_{\text{c}}}
\newcommand{\D}{\mathcal{D}}

\begin{document}

\preprint{RBRC-619}

\title{Model study of the sign problem in the mean-field approximation}
\author{Kenji Fukushima}
\affiliation{RIKEN BNL Research Center, Brookhaven National
 Laboratory, Upton, New York 11973, USA}
\author{Yoshimasa Hidaka}
\affiliation{RIKEN BNL Research Center, Brookhaven National
 Laboratory, Upton, New York 11973, USA}

\begin{abstract}
 We argue the sign problem of the fermion determinant at finite
 density.  It is unavoidable not only in Monte-Carlo simulations on
 the lattice but in the mean-field approximation as well.  A simple
 model deriving from Quantum Chromodynamics (QCD) in the double limit
 of large quark mass and large quark chemical potential exemplifies
 how the sign problem arises in the Polyakov loop dynamics at finite
 temperature and density.  In the color SU(2) case our mean-field
 estimate is in excellent agreement with the lattice simulation.  We
 combine the mean-field approximation with a simple phase reweighting
 technique to circumvent the complex action encountered in the color
 SU(3) case.  We also investigate the mean-field free energy, from the
 saddle-point of which we can estimate the expectation value of the
 Polyakov loop.
\end{abstract}


\maketitle


\section{Introduction}

  Quantum Chromodynamics (QCD) exhibits various states of matter
depending on the environment~\cite{review}:  A strongly coupled
quark-gluon plasma (sQGP) has been discovered at the Relativistic
Heavy Icon Collider (RHIC)~\cite{sQGP} and various color
superconductors presumably exist in the cores of the compact stellar
object~\cite{CSC}.  These are widely-known extreme states of
equilibrated QCD matter at high temperature and at high (baryon or
quark) density, respectively.  The study on sQGP properties at high
temperature has been led by the Monte-Carlo simulation of QCD on the
lattice~\cite{lattice}.  The lattice QCD results provide us with
fundamental information such as the phase transition temperature
$T_{\text{c}}$~\cite{Tc}, equation of state~\cite{Aoki:2005vt},
susceptibility~\cite{Bernard:2004je}, mesonic correlation above
$T_{\text{c}}$~\cite{charmonium}, etc.

  In contrast, at finite density, the lattice technique is not quite
successful so far;  it is hindered by the notorious problem that the
fermion determinant in the presence of nonzero quark chemical
potential $\mu_q$ is not positive semidefinite (i.e.\ not
nonnegative).  This problem is commonly refereed to as the fermion
sign problem~\cite{review:signproblem}.  The Monte-Carlo simulation
based on importance sampling requires positive semidefinite
probability for each gauge configuration.  In the presence of
$\mu_q\neq0$, however, the probability is no longer a well-defined
quantity due to a negative fermion determinant arising.  There have
been several techniques proposed to handle the sign problem, e.g.\ the
reweighting method~\cite{reweighting}, the Taylor expansion
method~\cite{Taylor}, the analytical continuation from an imaginary
chemical potential~\cite{imaginary}.  Although these proposals have
achieved partial successes when $\mu_q/T$ is small, no prescription
applicable at high density has been established yet.

  This work is aimed to point out that the sign problem is relevant
even in the mean-field treatment of gauge fields.  One can intuitively
understand it in the following way;  in the mean-field approximation
the partition function is estimated from the most dominant
contribution of particular configurations that have the largest
probability or the smallest free energy.  When the probability for
each gauge configuration is ill-defined due to the sign problem,
hence, the mean-field free energy may well be problematic.

  In the present work we shall focus on a specific manifestation of
the sign problem appearing in the simplest model at finite temperature
and density.  Because QCD is a highly nontrivial theory, applicability
of the mean-field approximation is quite limited.  We choose a finite
temperature model because the mean-field description presumably works
well to sketch the hot QCD phase transition.  The Polyakov loop plays
an essential role as an order parameter there~\cite{polyakov_loop}.
The dynamics of the Polyakov loop was closely examined some years ago
both in the lattice simulation~\cite{polyakov_lattice} and in the
mean-field approximation~\cite{weiss,strong}.  Recently, in addition,
the Polyakov loop dynamics near $T_{\text{c}}$ is specifically paid
attention~\cite{pl_model}.  There is also an interesting observation
that the entanglement between the chiral and Polyakov loop dynamics
turns out to be indispensable to understand the nature of the QCD
phase transitions~\cite{go_model,hatta,Fukushima:2003fw}.

  We have already known several indications from the mean-field
studies about the Polyakov loop behavior at $\mu_q\neq0$.  In the best
of our knowledge, the effective potential of the Polyakov loop at
finite temperature and density was first derived in
Ref.~\cite{KorthalsAltes:1999cp} in perturbative QCD in the one-loop
order.  (See also Ref.~\cite{Weiss:1987mp} for the potential with an
imaginary chemical potential.)  It is obvious from Eq.~(5) in
Ref.~\cite{KorthalsAltes:1999cp} that the Polyakov loop variable is
augmented to a complex valued variable when $\mu_q\neq0$, and thus the
effective potential turns complex.  It is quite nontrivial how to
derive meaningful information from such a complex effective potential;
the expectation value of the Polyakov loop cannot be fixed directly by
a minimum of the complex effective potential.  The same problem occurs
also in the chiral effective models with the Polyakov loop coupling
which was first formulated by one of the present authors in
Ref.~\cite{Fukushima:2003fw} and has been investigated
extensively~\cite{fnjl,Ratti:2005jh}, though the sign problem has been
almost overlooked.  Here, we would refer to a closely related work,
Ref.~\cite{Dumitru:2005ng}, in which the authors mentioned on the sign
problem in a matrix model of the Polyakov loop.  It should be noted
that the mean-field free energy as seen in
Refs.~\cite{Ratti:2005jh,Dumitru:2005ng} is not complex unlike the
effective potential in Ref.~\cite{KorthalsAltes:1999cp} so that one
can determine the expectation value of the Polyakov loop without
ambiguity.  We will go into details of this issue later.  Our findings
are all consistent with what has been foreseen in
Ref.~\cite{Dumitru:2005ng}.

  The concept of this work lies not in solving the sign problem but in
observing the sign problem in a manageable way, so to speak, in a
clean environment.  We also demonstrate that a technique which we call
the \textit{phase reweighting method} works fine as a practical
prescription in the similar but not the same spirit as in the finite
density lattice simulation.  We will employ the dense-heavy model
obtained from QCD in the double limit of large quark mass and large
quark chemical potential~\cite{Blum:1995cb} (see also discussions in
Ref.~\cite{Bender:1991gn}).  The reasons we adopt the dense-heavy
model are as follows:  First, the fermion determinant is exactly
calculable as a function of the gauge field.  Second, neither the
chiral condensate nor the diquark condensate is involved in the
dynamics owing to large quark mass.  Third, lattice data is available
from Ref.~\cite{Blum:1995cb}.  From these three reasons the
dense-heavy model is considered to be an appropriate implement for our
purpose to scrutinize the sign problem within the mean-field
approximation.

  This paper is organized as follows.  In Sec.~\ref{sec:sign_problem}
we make a brief overview on the sign problem.  We formulate the model
in Sec.~\ref{sec:dense_heavy} and the mean-field approximation in
Sec.~\ref{sec:mean_field}.  Then, in Sec.~\ref{sec:su2}, we examine
the color SU(2) case first which is free from the sign problem in
order to make sure if the mean-field approximation is reasonable.  We
next proceed to the SU(3) calculation.  In Sec.~\ref{sec:su3} we
explain the phase reweighting method to circumvent the SU(3) complex
fermion determinant and then discuss the validity of our method by
viewing the mean-filed effective potential in
Sec.~\ref{sec:MFFreeEnergy}.  Section~\ref{sec:summary} is devoted to
the summary.


\section{Sign Problem}
\label{sec:sign_problem}

  Here are general discussions on the sign problem of the fermion
determinant at finite density.  Readers who are already familiar with
it can skip to the model study starting from Sec.~\ref{sec:model}.  We
will later demonstrate how the dense-heavy model concretely embodies
the general features mentioned in this section.

  The fermion determinant in Euclidean space-time with a quark mass
$m_q$ and a quark chemical potential $\mu_q$ takes the form of
\begin{equation}
 \det\mathcal{M}(\mu_q)\equiv
  \det\bigl[\gamma_\mu D^\mu + \gamma_4\mu_q +m_q\bigr] \,,
\end{equation}
where $D^\mu\equiv\partial^\mu-igA^\mu$ is the covariant derivative.
It is a well-known argument that
$\det\mathcal{M}(\mu_q)=\det\gamma_5\mathcal{M}(\mu_q)\gamma_5
=\{\det\mathcal{M}(-\mu_q)\}^\ast$ and thus the fermion determinant is
real in the zero density ($\mu_q=0$) case.  Alternatively, we can
explicitly look into the eigenvalue spectrum of the Dirac operator to
confirm that the determinant is positive semidefinite as follows:  For
$\mu_q=0$, when $\psi_n$ is an eigenstate of $\gamma_\mu D^\mu$, the
eigenvalue $\lambda_n$ is pure imaginary because $\gamma_\mu D^\mu$ is
anti-Hermitian where $\gamma_\mu$'s are Hermitian in our convention.
Since the mass term is simply proportional to unity, $\psi_n$ is also
an eigenstate of $\gamma_\mu D^\mu+m_q$ with the eigenvalue
$\lambda_n+m_q$.  One can show that $\gamma_5 \psi_n$ is an eigenstate
of $\gamma_\mu D^\mu+m_q$ having the eigenvalue
$-\lambda_n+m_q=(\lambda_n+m_q)^\ast$ from the property
$\gamma_5\gamma_\mu D^\mu\gamma_5=-\gamma_\mu D^\mu$.  Therefore, the
determinant consists of a pair of $\lambda_n+m_q$ and
$(\lambda_n+m_q)^\ast$ which makes the whole determinant real and
nonnegative as $|\lambda_n+m_q|^2\geq0$.

  In the presence of the chemical potential term the fermion
determinant is not necessarily positive semidefinite.  When $\psi_n$
is an eigenstate of $\gamma_\mu D^\mu+\gamma_4\mu_q$ with the
eigenvalue $\lambda_n$, in the same was as above, one can show that
$\gamma_5\psi_n$ is also an eigenstate of
$\gamma_\mu D^\mu+\gamma_4\mu_q$ with the eigenvalue $-\lambda_n$
which is not $\lambda_n^\ast$ because $\gamma_\mu D^\mu+\gamma_4\mu_q$
is no longer anti-Hermitian.  It is apparent from this explanation
that, if $\mu_q$ is pure imaginary~\cite{imaginary},
$\gamma_\mu D^\mu+\gamma_4\mu_q$ is anti-Hermitian so that the fermion
determinant turns nonnegative.

  If a considered theory has degenerate fermions associated with an
internal symmetry under the transformation $T$ and the chemical
potential $\mu_q$ is replaced by a matrix $\boldsymbol{\mu}$ in
internal space transforming like
\begin{equation}
 T^{-1}\,\boldsymbol{\mu}\,T=-\boldsymbol{\mu}\;,
\label{eq:change}
\end{equation}
then
$\det\mathcal{M}(\boldsymbol{\mu})
=\{\det\mathcal{M}(\boldsymbol{\mu})\}^\ast$, so that the determinant
becomes real.

  A well-known realization of this comes from the isospin degrees of
freedom in flavor ($u$,$d$) space.  The isospin chemical potential
$\mu\propto\tau_3$ and $T=i\tau_2$ (or $T=\tau_1$) certainly satisfies
Eq.~(\ref{eq:change}) where $\tau$'s are the Pauli matrices in
($u$,$d$) space~\cite{Son:2000xc}.

  Another realization is the color SU(2) case~\cite{su2} that is
relevant to our model study as we will argue later.  The
$C$-transformation changes the quark chemical potential $\mu_q$ as in
Eq.~(\ref{eq:change}).  In general cases, however, the
$C$-transformation does not give an escape from the sign problem
because it also swaps color and anticolor.  That is,
\begin{equation}
 C^{-1}\gamma_5\,\gamma_\mu D^\mu\,\gamma_5 C
  = \bigl[\gamma_\mu\bigl(\partial^\mu-ig(A^\mu)^C\bigr)\bigr]^\ast
\end{equation}
with $(A^\mu)^C\equiv-(A^\mu)^\ast$.  Special for $\Nc=2$ is that
anticolor is not distinguishable from color since two doublets can
make a singlet.  Actually $\sigma_2(A^\mu)^C\sigma_2=A^\mu$ where
$\sigma$'s are the Pauli matrices in color space.  Therefore the SU(2)
case is free from the sign problem.

  We should note that the fermion determinant being complex is not
necessarily harmful on its own.  Rather, it is important whether the
real or imaginary part of the fermion determinant is positive
semidefinite or not.  Even though the determinant evaluated for a
certain $A^\mu$ is a complex number, the functional integral over
$A^\mu$ amounts to a real value for physical observables.  It is
understood in view of the relation,
\begin{equation}
 \begin{split}
 & \det\bigl[\gamma_\mu\bigl(\partial^\mu-ig(A^\mu)^C\bigr)
  +\gamma_4\mu_q+m_q\bigr] \\
 =& \Bigl\{\det\bigl[\gamma_\mu D^\mu+\gamma_4\mu_q
  +m_q\bigr]\Bigr\}^\ast \,.
 \end{split}
\label{eq:cconj}
\end{equation}
We see clearly that the real (imaginary) part of the determinant is
$C$-even ($C$-odd).  For a $C$-even ($C$-odd) observable, thus, the
imaginary (real) part of the determinant vanishes after integration
over $A^\mu$.  Accordingly the genuine problem stems from that the
real or imaginary part of the determinant may change its sign
depending on the configuration $A^\mu$.


\section{Model Study}
\label{sec:model}

  We will analyze a simple model to see the sign problem occurring in
the mean-field level.  As a practice to study the model, we will make
the mean-field approximation in the SU(2) case for which we do not
have to face the sign problem.  We will then observe the sign problem
in the SU(3) calculation and attempt the phase reweighting method to
deal with the complex phase of the fermion determinant.


\subsection{Dense-Heavy Model}
\label{sec:dense_heavy}

  We will closely analyze a lattice model with dense heavy
quarks~\cite{Blum:1995cb}.  Let us consider QCD in the limit of
$\mu_q\to\infty$ so that we can drop antiquarks.  We shall
simultaneously take another limit of $m_q\to\infty$ which renders all
quarks static.  Under such limits we can evaluate the staggered
fermion determinant exactly to reach the fermion action,
\begin{align}
 e^{-\Sf[L]} \equiv & \det\bigl[\gamma_\mu D^\mu+\gamma_4\mu_q
  +m_q\bigr]\notag\\
 \to & \bigl[\det(1+\epsilon L)\bigr]^{\Nf/4} \,,
\end{align}
where $\Nf$ is the number of flavors.  Not to go into subtlety of the
flavor counting inherent to the staggered formalism, which is not of
our interest, we set $\Nf=4$ throughout this paper.

  We take those two limits in a way characterized by the parameter
$\epsilon$ ranging from zero to infinity;
\begin{equation}
 \epsilon\equiv\Bigl(\frac{e^{\mu_q a}}{2m_q a}\Bigr)^{N_\tau}\,.
\end{equation}
We will often call this model parameter as the ``density'' parameter
because $\epsilon$ has strong correlation to the quark number density
as seen in Fig.~\ref{fig:edep_density_su2} for the SU(2) case and
Fig.~\ref{fig:edep_density_su3} for the SU(3) case.  Here $N_\tau$ is
the number of the lattice sites in the temporal direction, that is,
the inverse temperature $1/T$ is given by $N_\tau a$ with the lattice
spacing $a$.  In this model quarks are allowed to propagate only in
the positive temporal direction.  Each time a quark with mass $m_q$
travels by one temporal lattice, it picks up the hopping parameter
$1/(2m_q a)$ and the gauge invariant chemical potential factor
$e^{\mu_q a}$~\cite{Hasenfratz:1984em}.  After $N_\tau$ hops, a quark
winds around the temporal circle.  It results in the weight $\epsilon$
for the quark excitation represented by the Polyakov loop $L$ in the
fundamental representation,
\begin{align}
 L(\vec{x}) &\equiv \prod_{x_4=a}^{N_\tau a}U_4(\vec{x},x_4) \notag\\
 &\equiv\mathcal{P}\exp\biggl[\,ig\!\int_0^{1/T}\!\! dx_4\,
  A_4(\vec{x},x_4)\biggr] \,.
\end{align}
The first line is the expression of the Polyakov loop in terms of link
variables on the lattice and second is in the continuum.  It does not
matter whichever expression we use since the SU($\Nc$) matrix $L$
plays the role of the dynamical variable in our model and we will not
return to its definition.

  The calculation of the determinant in color space is explicitly
doable.  After all we have the fermion action,
\begin{equation}
 e^{-\Sf[L]} = \prod_{\vec{x}}\bigl[1+\epsilon^2
  +2\epsilon\ell\bigr]
\label{eq:sf_su2}
\end{equation}
in the color $\Nc=2$ case (and $\Nf=4$ is implicit as we already
noted) and
\begin{equation}
 e^{-\Sf[L]} = \prod_{\vec{x}}\bigl[1+\epsilon^3
  +3\epsilon\ell+3\epsilon^2\ell^\ast\bigr]
\label{eq:sf_su3}
\end{equation}
in the color $\Nc=3$ case.  In the above expressions we employed the
traced Polyakov loop defined as
\begin{equation}
 \ell = \frac{1}{\Nc}\tr L,
\end{equation}
where $\tr$ is taken in fundamental color space.   We remark that the
$C$-transformation changes $\ell$ to $\ell^\ast$ and vice versa, where
$\ell$ is real for $\Nc=2$ and generally complex for $\Nc\neq2$.

  It is obvious that the $\Nc=2$ determinant (\ref{eq:sf_su2}) is real
and positive semidefinite for any $\epsilon$ because
$1+\epsilon^2+2\epsilon\ell\ge(1-|\epsilon|)^2\ge0$, while the $\Nc=3$
expression (\ref{eq:sf_su3}) suffers the sign problem for nonzero
$\epsilon$;  the determinant can be complex except when either
$\epsilon=0$ (zero density), $\epsilon=1$ (half-filling), or
$\epsilon\to\infty$ (full-filling).  We can intuitively understand why
the SU(3) determinant becomes real for $\epsilon=1$.  One-quark
excitation represented by $\ell$ and two-quark excitation that is
equivalent to one-antiquark excitation represented by $\ell^\ast$
occur with the common weight $\epsilon=\epsilon^2=1$ because of
half-filling (see $n=0.5$ at $\epsilon=1$ in
Figs.~\ref{fig:edep_density_su2} and \ref{fig:edep_density_su3}).
Therefore, in effect, the system has equality in number of quarks and
(effective) antiquarks just like in the zero density case.  It is not
an escape from the sign problem, however.  In the $\Nc=3$ case
$\ell+\ell^\ast$ can take a value ranging from $-1$ to $2$, and thus
the determinant at $\epsilon=1$, i.e.\
$\det(1+L)\propto 2+3(\ell+\ell^\ast)$ is real but can be negative for
$-1<\ell+\ell^\ast<-2/3$.

  We note here one more important feature of the model.  The fermion
determinant is invariant under the duality
transformation~\cite{Blum:1995cb},
\begin{equation}
 \epsilon \:\leftrightarrow\: 1/\epsilon \quad\text{and}\quad
 \ell \:\leftrightarrow\: \ell^\ast \,,
\end{equation}
by which it is sufficient for us to investigate the model in the
region $\epsilon\in[0,1]$ and the outer region $\epsilon\in(1,\infty]$
can be deduced by means of the duality.

  Regarding the gluodynamics, we assume the nearest neighbor
interaction between the Polyakov loops,
\begin{equation}
 \Sg[L]=-\Nc^2 J\sum_{\text{n.n.}} \ell(\vec{x})\,
  \ell^\ast(\vec{y}).
\label{eq:sg}
\end{equation}
This action looks pretty simple and still reproduces the fundamental
nature of the phase transition in the pure gluonic sector;  the
action~(\ref{eq:sg}) leads to a second-order phase transition for the
SU(2) case and a first-order phase transition for the SU(3)
case~\cite{Kogut:1981ez}, which is in agreement with the lattice QCD
simulation and the theoretical expectation from center
symmetry~\cite{polyakov_loop}.

  One can interpret $J$ as a model parameter specifying the
``temperature'' of the system.  In the strong coupling expansion, in
fact, $J$ is related to $T$ through $J=\exp[-\sigma a/T]$ where
$\sigma$ is the string tension.  In this work we shall leave $J$ as a
model parameter as it is, for we do not want to introduce any further
modeling into our analyses.  In other words, our aim is to form a
model not to imitate QCD itself but to mimic the QCD sign problem.

  The effective action that defines our model is eventually given by
\begin{equation}
 S[L]=\Sg[L]+\Sf[L]
\end{equation}
with the ``density'' parameter $\epsilon$ contained in $\Sf[L]$ and
the ``temperature'' parameter $J$ in $\Sg[L]$.


\subsection{Mean Field Approximation}
\label{sec:mean_field}

  In the finite-temperature field theory the free energy is evaluated
in the functional integral form like the effective action as
\begin{equation}
 e^{-f\cdot V/T}=\int\!\D L\,e^{-S[L]} \;.
\end{equation}
We make use of the mean-field technique to approximate the free
energy.  Our ansatz for the mean-field action
is~\cite{go_model}
\begin{equation}
 \Smf[L] \equiv -\frac{x}{2}\sum_{\vec{x}}\bigl[\ell(\vec{x})
  +\ell^\ast(\vec{x})\bigr] \;.
\label{eq:ansatz}
\end{equation}
In case of pure gluonic theories, the mean-field $x$ is simply
proportional to the Polyakov loop expectation value;
$x=12\Nc^2 J\langle\ell\rangle$.  In the presence
of fermionic contributions, however, there is no simple relation
between them.  Besides, $\langle\ell\rangle$ and
$\langle\ell^\ast\rangle$ in the SU(3) case have different dependence
on $\mu_q$.  One might think that there should be two independent
mean-fields to deal with differing $\langle\ell\rangle$ and
$\langle\ell^\ast\rangle$.  This idea has much to do with the sign
problem actually, and we will come back to this point later.

Then the mean-field free energy can be estimated as
\begin{equation}
 \fmf(x)\cdot V/T = \langle S[L]-\Smf[L]
  \rangle_{\text{mf}} -\ln\!\int\D L\,e^{-\Smf[L]},
\label{eq:free_energy}
\end{equation}
where the average $\langle\cdots\rangle_{\text{mf}}$ is taken by the
mean-field action $\Smf[L]$.  Roughly speaking, the first part
corresponds to the internal energy and the logarithmic part is the
entropy.  We fix $x$ so as to minimize $\fmf(x)$.  Once $\Smf[L]$ is
known with $x$ determined, the expectation value of any observable
$\mathcal{O}[L]$ as a function of the Polyakov loop can be estimated
by the group integration over $L$ with the mean-field action
$\Smf[L]$;
\begin{equation}
 \langle\mathcal{O}[L]\rangle \simeq \langle\mathcal{O}[L]
  \rangle_{\text{mf}} \equiv \frac{\displaystyle\int\!dL\,
  \mathcal{O}[L]\,e^{-\Smf[L]}}{\displaystyle \int\!dL\,
  e^{-\Smf[L]}} \,.
\end{equation}
For instance, the quark number density per color degrees of freedom is
available by calculating
\begin{equation}
 n \equiv -\frac{1}{\Nc}\cdot\frac{\partial f}
  {\partial\mu_q}\simeq\frac{\epsilon}{\Nc V}\biggl\langle
  \frac{d\Sf}{d\epsilon}\biggr\rangle_{\text{mf}} \,.
\label{eq:density}
\end{equation}
One can calculate the Polyakov loop susceptibility $\chi$ in the same
way which reflects information of the deconfinement phase transition.
If one directly uses $\mathcal{O}[L]=\ell^2$ in the mean-field
approximation, however, nothing becomes singular at the critical point
unless the fluctuation of $\ell$ is taken into account.  Equivalently
one can estimate the susceptibility from the inverse curvature of the
effective potential because the one-loop fluctuation leads to the
trace of propagator for $\langle\ell^2\rangle$, that is, the inverse
screening mass.  Since $\langle\ell\rangle$ and
$\langle\ell^\ast\rangle$ are uniquely determined given $x$ is fixed
by the free energy~(\ref{eq:free_energy}), one can regard
$\fmf(x)$ as a function of $\langle\ell\rangle$ or
$\langle\ell^\ast\rangle$.  The Polyakov loop susceptibility is then
\begin{equation}
 \biggl(\frac{\partial^2 \fmf}{\partial\langle\ell\rangle^2}\biggr)^{-1}
  =\biggl(\frac{\partial\langle\ell\rangle}{\partial x}\biggr)^2
  \biggl(\frac{\partial^2 \fmf}{\partial x^2}\biggr)^{-1} \,,
\end{equation}
where we used $\partial\fmf/\partial x=0$.  Of course, the
susceptibility defined in terms of $\ell^\ast$ is available with
$(\partial\langle\ell\rangle/\partial x)^2$ replaced by
$(\partial\langle\ell^\ast\rangle/\partial x)^2$.  The difference thus
lies only in the nonsingular coefficient we are not interested in.
For our purpose it is rather convenient to focus on the singular part
alone discarding the difference of $\langle\ell\rangle$ and
$\langle\ell^\ast\rangle$.  Hence, we define the susceptibility as
\begin{equation}
 \chi \equiv \biggl(\frac{\partial^2 \fmf}{\partial^2 x}
  \biggr)^{-1}
\label{eq:sus}
\end{equation}
for presenting our numerical results.

  Now that we finish explaining our approximations and computational
procedures, let us step forward to the model analysis.


\subsection{SU(2) Results}
\label{sec:su2}

  We consider the model in the color SU(2) case first, as we
mentioned, to see how the mean-field approximation works apart from
the sign problem.

\begin{figure}
 \includegraphics[width=7.5cm]{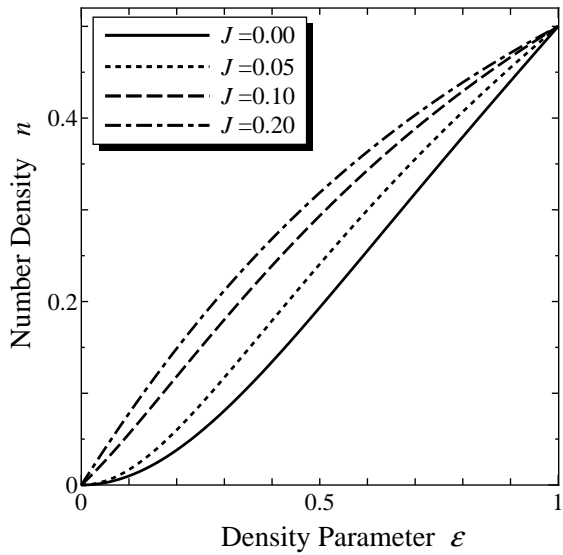}
 \caption{Correlation between the density parameter $\epsilon$ and the
 quark number density per lattice site (i.e.\ density in the unit of
 the lattice spacing) divided by $\Nc=2$.  It is obvious from the
 figure that $\epsilon=0$ is the zero-density ($n=0$) state and
 $\epsilon=1$ is the half-filling ($n=0.5$) one.}
 \label{fig:edep_density_su2}
\end{figure}

\begin{figure}
 \includegraphics[width=7.5cm]{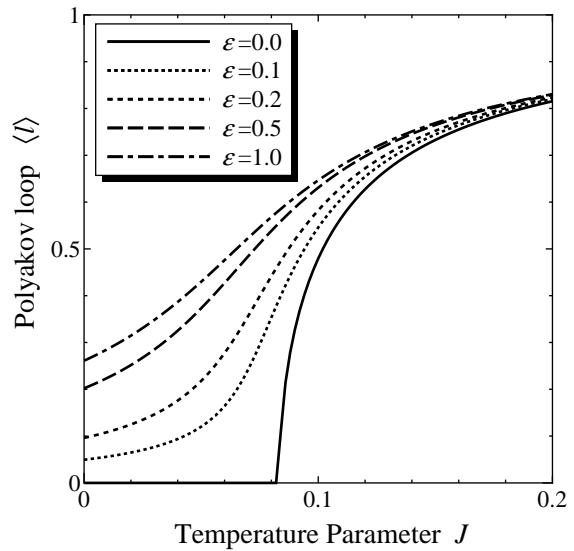}
 \caption{SU(2) (traced) Polyakov loop $\langle\ell\rangle$ as a
 function of the coupling $J$ at various density parameters;
 $\epsilon=$0 (solid), 0.1 (dotted), 0.2 (short-dashed), 0.5 (dashed),
 and 1.0 (dotted-dashed).  The second-order phase transition at
 $\epsilon=0$ occurs at $J_{\text{c}}\simeq0.083$.  The transitional
 behavior is gradually smeared by the center symmetry breaking terms
 in the fermion determinant as $\epsilon$ grows larger.}
 \label{fig:Jdep_su2}
\end{figure}

\begin{figure}
 \includegraphics[width=7.5cm]{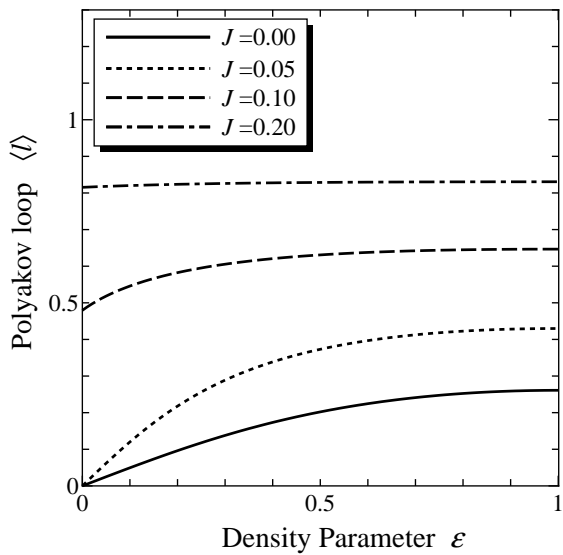}
 \caption{SU(2) Polyakov loop $\langle\ell\rangle$ as a function of
 the density parameter $\epsilon$ at various temperature parameters;
 $J$=0 (solid), 0.05 (short-dashed), 0.1 (dashed), and 0.2
 (dotted-dashed).  The Polyakov loop becomes insensitive to $\epsilon$
 as $J$ goes larger, which is consistent with Fig.~\ref{fig:Jdep_su2}
 in which the results at various $\epsilon$ converge at large $J$.}
 \label{fig:edep_su2}
\end{figure}

  Figure~\ref{fig:edep_density_su2} shows the results for the quark
number density $n$ as a function of the density parameter $\epsilon$
using Eq.~(\ref{eq:density}).  For $\epsilon>1$ the duality relation
$n(\epsilon)=1-n(1/\epsilon)$ enables us to deduce the number density.
We can immediately confirm from this plot and the duality relation
that the density parameter specifies the quark number density uniquely
which monotonously approaches unity as $\epsilon\to\infty$.  It should
be noted that the half-filling $n=0.5$ realizes at $\epsilon=1$ and
there is no $J$ dependence at all then.

  Let us look at the phase transition seen in the Polyakov loop
behavior with increasing $J$.  The deconfinement phase transition in
the SU(2) pure gluonic theory is known to be second-order belonging to
the same universality class as the Ising
model~\cite{polyakov_loop,polyakov_lattice,Engels:1989fz}.
In our model at $\epsilon=0$ we have a continuous transition at
$J=J_{\text{c}}\simeq0.083$ as indicated by the solid curve in
Fig.~\ref{fig:Jdep_su2}.  The presence of dynamical quarks acts on the
Polyakov loop variable as an external field breaking center symmetry.
In fact, the results at nonzero $\epsilon$ in Fig.~\ref{fig:Jdep_su2}
are not of transition but of crossover.

  We plot the density dependence of the Polyakov loop behavior in
Fig.~\ref{fig:edep_su2}.  The density effects generally tend to make
the Polyakov loop larger, and eventually, the Polyakov loop becomes
insensitive to the density in the large $J$ (i.e.\ high temperature)
region.  It is because both the temperature and the density break
center symmetry spontaneously and explicitly, respectively, having the
Polyakov loop saturated.  Therefore, the $J$ dependence is less for
larger $\epsilon$ and the $\epsilon$ dependence is less for larger
$J$.

\begin{figure}
 \includegraphics[width=7.5cm]{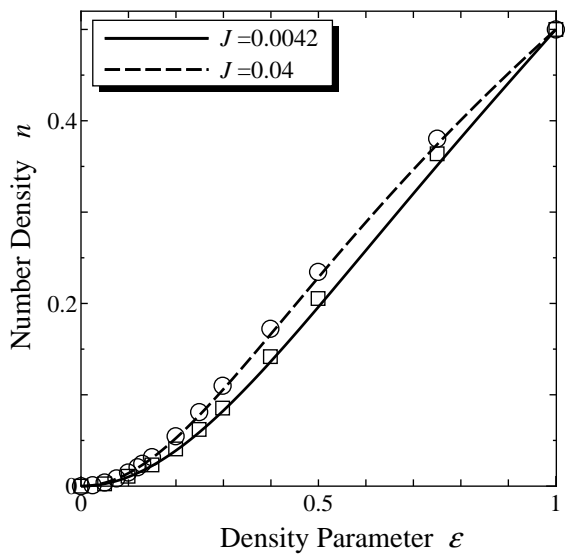}
 \caption{Comparison to the number density measured on the lattice at
 $4/g^2=2.0$ (circle) and $4/g^2=1.5$ (square) taken from Fig.~1 in
 Ref.~\cite{Blum:1995cb}.}
 \label{fig:Tom_fit_num}
\end{figure}

\begin{figure}
 \includegraphics[width=7.5cm]{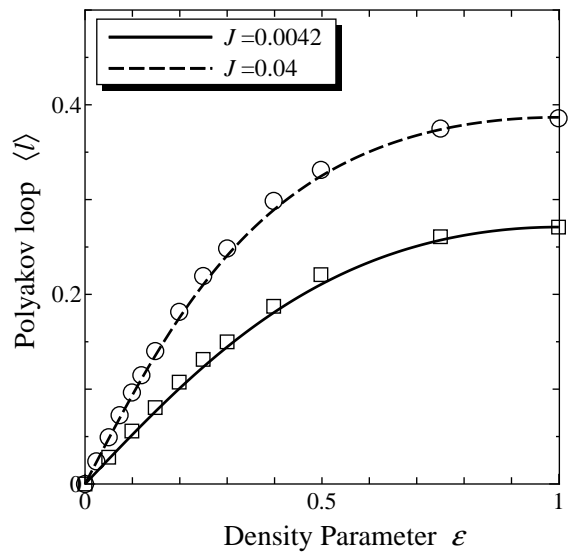}
 \caption{Comparison to the SU(2) Polyakov loop measured on the
 lattice at $4/g^2=2.0$ (circle) and $4/g^2=1.5$ (square) taken from
 Fig.~2 in Ref.~\cite{Blum:1995cb}.}
 \label{fig:Tom_fit_P}
\end{figure}

  Our mean-field outputs are to be compared with the lattice
simulation in Ref.~\cite{Blum:1995cb};  our
Figs.~\ref{fig:edep_density_su2} and \ref{fig:edep_su2} correspond to
Figs.~1 and 2 presented in Ref.~\cite{Blum:1995cb}, respectively.  We
cannot expect a quantitative coincidence because our ansatz for the
pure gluonic action $\Sg[L]$ is only a crude approximation of QCD and
besides we neglect the renormalization of the Polyakov loop in the
mean-field treatment.  Nevertheless, the agreement turns out to be
surprisingly good beyond our expectation if we treat the model
parameter $J$ as a fitting parameter incorporating the undetermined
effect of the Polyakov loop renormalization, which is implied by the
ansatz (\ref{eq:sg}).  In such a way, we can fix $J=0.0042$ and
$J=0.04$ to reproduce the SU(2) Polyakov loop \textit{only} at
$\epsilon=1$ for $4/g^2=2.0$ and $4/g^2=1.5$, respectively.  We would
emphasize that we did \textit{not} use the data of the Polyakov loop
at $\epsilon\neq1$ and not the data of the number density at all.
Nevertheless, as clearly seen from the comparisons in
Figs.~\ref{fig:Tom_fit_num} and \ref{fig:Tom_fit_P}, our numerical
results fit \textit{all} of the lattice data pretty well.  We can
conclude from this observation that the main QCD corrections to our
ansatz (\ref{eq:sg}) could be absorbed into the renormalization of the
coupling alone.  We are now confident that the mean-field treatment is
a fairly acceptable approximation for this type of problem.

\begin{figure}
 \includegraphics[width=7.5cm]{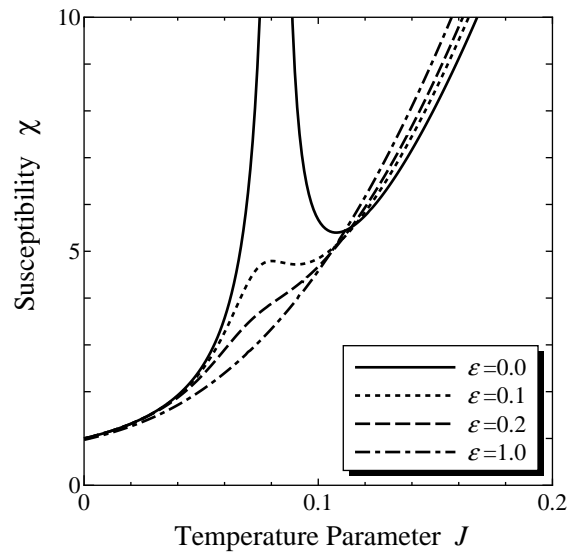}
 \caption{Susceptibility relevant to the SU(2) Polyakov loop $\ell$ as
 a function of $J$ at various density parameters; $\epsilon=$0.0
 (solid), 0.1 (short-dashed), 0.2 (dashed), and 1.0 (dotted-dashed).
 The $\epsilon=0$ result has a divergence at $J=J_{\text{c}}$
 characteristic to the second-order phase transition.}
 \label{fig:Jdep_sus_su2}
\end{figure}

  Finally let us check that the susceptibility (\ref{eq:sus}) diverges
at $\epsilon=0$ and $J=J_{\text{c}}$.  Figure~\ref{fig:Jdep_sus_su2}
shows the susceptibility as a function of $J$ at various $\epsilon$.  In
the plot $\chi$ becomes greater with increasing $J$ because we did not
include $\partial\langle\ell\rangle/\partial x$ in our definition of
$\chi$ in Eq.~(\ref{eq:sus}).  It is intriguing to remark that the
$\epsilon=0.1$ result is not really critical in view of $\chi$, while
the crossover at $\epsilon=0.1$ in Fig.~\ref{fig:Jdep_su2} looks
rather close to a phase transition.  Actually $\chi$ is a more
informative quantity to judge how critical the crossover is in fact.

\begin{figure*}
 \includegraphics[width=14cm]{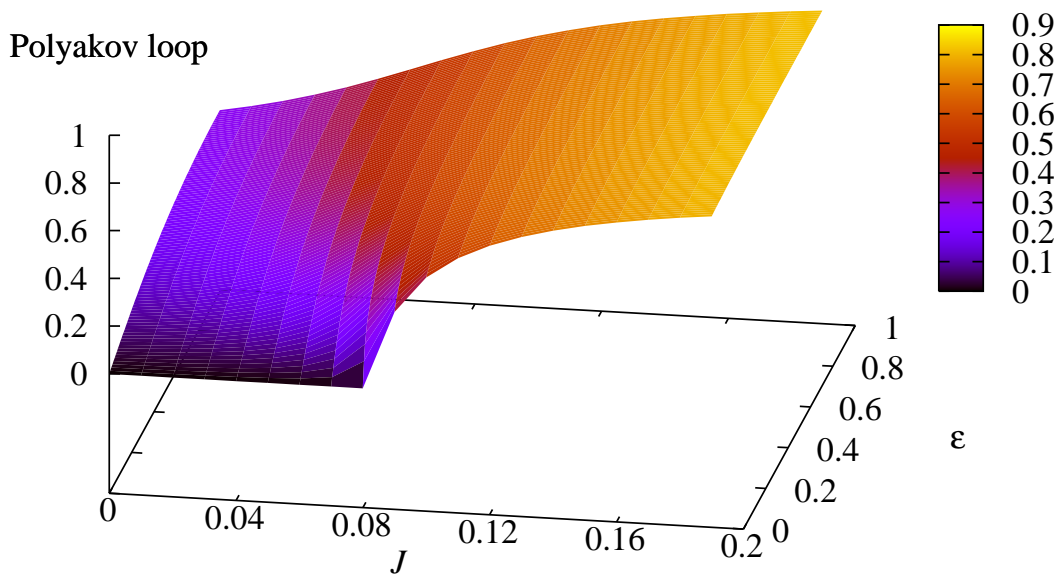}
 \caption{Three-dimensional plot of the fundamental Polyakov loop in
 the SU(2) case as a function of the temperature parameter $J$ and the
 density parameter $\epsilon$.}
 \label{fig:3d_su2}
\end{figure*}

  In summary of the SU(2) Polyakov loop dynamics at finite temperature
and density, we shall depict a three-dimensional plot of
$\langle\ell\rangle$ in Fig.~\ref{fig:3d_su2} as a function of the
``temperature'' $J$ and the ``density'' $\epsilon$.  We immediately
see general tendency that $\langle\ell\rangle$ grows up with
increasing $J$ and $\epsilon$.

\begin{figure}
 \includegraphics[width=7.5cm]{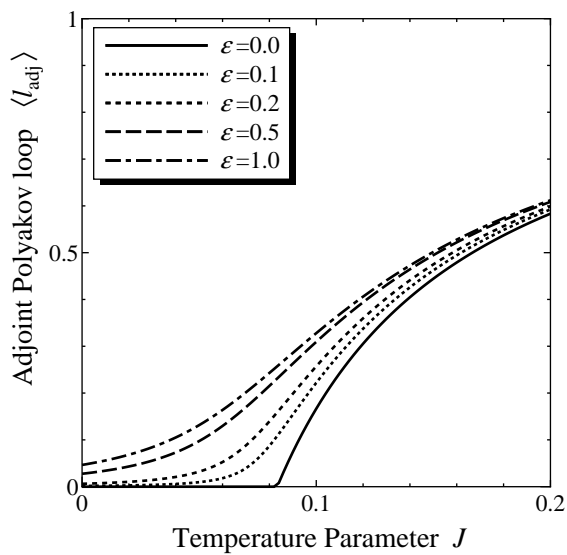}
 \caption{Dependence of the SU(2) adjoint Polyakov loop
 $\langle\ell_{\text{adj}}\rangle$ on the temperature parameter $J$ at
 various density parameters; $\epsilon$=0.0 (solid), 0.1 (dotted), 0.2
 (short-dashed), 0.5 (dashed), and 1.0 (dotted-dashed).}
 \label{fig:Jdep_ad_su2}
\end{figure}

  Before closing this subsection we will comment a bit on the adjoint
Polyakov loop whose definition is
\begin{equation}
 \ell_{\text{adj}} \equiv \frac{1}{\Nc^2-1}\tr L^{\text{adj}}
  =\frac{1}{\Nc^2-1}\Bigl(\Nc^2|\ell|^2-1\Bigr) \;.
\label{eq:ad_pol}
\end{equation}
The adjoint Polyakov loop is no longer an order parameter because the
adjoint representation is not faithful by the center of the gauge
group.  This quantity has been closely argued in
Ref.~\cite{Dumitru:2003hp} and there still remain subtleties.  Not in
the SU(2) but in the SU(3) case, it has been found that
$\langle\ell_{\text{adj}}\rangle$ takes an infinitesimally small value
in the confined phase.  One possible explanation would be that the
group integration over $L$ is important in the center symmetric regime
where nonperturbative phenomena like color confinement are relevant
(see also Ref.~\cite{Lenz:1998qk}).  As a matter of fact, the
SU($\Nc$) group integration over $\ell_{\text{adj}}$ turns out to be
zero, if we are allowed to disregard the dynamics.  Even with the
dynamics taken into account within the mean-field approximation, as
shown in Fig.~\ref{fig:Jdep_ad_su2}, the argument should hold so
$\langle\ell_{\text{adj}}\rangle\simeq 0$ in the low temperature side.
However, the adjoint Polyakov loop behavior in our study should be
understood only up to a qualitative level.  It is pointed out in
Ref.~\cite{Dumitru:2003hp} that the renormalization for
$\ell_{\text{adj}}$ is significant, which is not considered in our
present treatment.


\subsection{SU(3) case}
\label{sec:su3}

  In case of $\Nc=3$ we have to tackle the sign problem.  We are not
capable of solving QCD exactly like the lattice simulation, so one
might think that within the framework of approximations one is allowed
to impose the mean-field ansatz (\ref{eq:ansatz}) to get some results
anyhow.  The free energy after the integration over $L$ could be real
in terms of $x$.  It seems to work at least as a rough estimate that
is worth trying first.

  The serious flaw in such a simple strategy is that
$\langle\ell\rangle=\langle\ell^\ast\rangle$ is inevitably concluded.
It is, however, contradict to the lattice results~\cite{Taylor} and
the model analyses~\cite{Dumitru:2005ng} where
$\langle\ell\rangle\neq\langle\ell^\ast\rangle$ has been observed at
finite density.  If the mean-field ansatz (\ref{eq:ansatz}) is
extended to having two variables $x$ and $y$ in order to take account
of the difference between $\langle\ell\rangle$ and
$\langle\ell^\ast\rangle$, the price to pay is that the mean-field
free energy is not convex.  We will revisit this issue latter.  In any
case, though the appearance might be unalike, the difficulty of the
sign problem is conserved even in the mean-field approximation unless
the difference $\langle\ell\rangle\neq\langle\ell^\ast\rangle$
is neglected.

  In our work we shall elucidate that the ``phase reweighting method''
is one way to resolve these difficulties.  We should, however, note
that the reweighting method in the present context is one
\textit{approximation} scheme unlike in the lattice simulation.  That
is, the reweighting method is expected to be precise if the number of
configurations is infinitely large, and thus the lattice simulation
with infinite number of configurations generated could provide us with
the exact answer in principle, while the mean-field approximation
picking up only the most dominant configuration cannot.


\subsubsection{Method}

  The point of the method is that we decompose the fermion determinant
into one part that gives the positive semidefinite probability and the
other part that is regarded as the observable whose average is taken
by configurations.

  The complex fermion determinant consists of the $C$-even magnitude
and the $C$-odd phase.  Accordingly, the fermion action can be
rewritten as
\begin{equation}
 \Sf[L]= \Smag[L]+i\Theta[L] \,,
\end{equation}
where
\begin{align}
 \Smag[L] &= -\sum_{\vec{x}} \ln\bigl|1+\epsilon^3
  +3\epsilon\ell+3\epsilon^2\ell^\ast\bigr| \,, \\
 \Theta[L] &= -\sum_{\vec{x}}\arg\bigl(1+\epsilon^3+3\epsilon\ell
  +3\epsilon^2\ell^\ast\bigr) \,.
\label{eq:phase}
\end{align}
With these definitions we approximate the expectation value of
$\mathcal{O}[L]$ by the one obtained as follows;
\begin{equation}
 \langle\mathcal{O}[L]\rangle\simeq \frac{\displaystyle
  \bigl\langle\mathcal{O}[L]\,e^{-i\Theta[L]}\bigr\rangle_{\text{mf}}}
  {\displaystyle \bigl\langle e^{-i\Theta[L]}\bigr\rangle_{\text{mf}}}
  \;.
\label{eq:rew}
\end{equation}
Here $\Smf[L]$ or $x$ is fixed from the free energy with the action
$\Sg[L]+\Smag[L]$, so that $x$ encompasses the information of
$\Smag[L]$ implicitly.  This scheme is the same as what has been
adopted in the lattice simulation in Ref.~\cite{Blum:1995cb}.

  Here we would draw attention to a related work;  in
Ref.~\cite{deForcrand:1999cy} the correlation between $\text{Im}\ell$
and $\Theta[L]$ was investigated in the lattice simulation, which is
apparent in our case from the expression (\ref{eq:phase}).


\subsubsection{Results}
  
\begin{figure}
 \includegraphics[width=7.5cm]{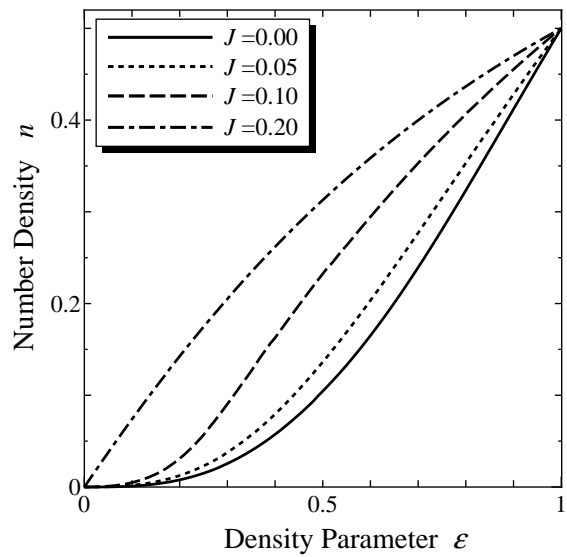}
 \caption{Correlation between the density parameter $\epsilon$ and the
 quark number density $n$ per lattice site divided by $\Nc=3$.  The
 gross feature is similar to Fig.~\ref{fig:edep_density_su2};
 $\epsilon=0$ is the zero density state ($n=0$) and $\epsilon=1$ is
 the half-filling state ($n=0.5$) in this SU(3) case as well as in the
 SU(2) case.}
 \label{fig:edep_density_su3}
\end{figure}

\begin{figure}
 \includegraphics[width=7.5cm]{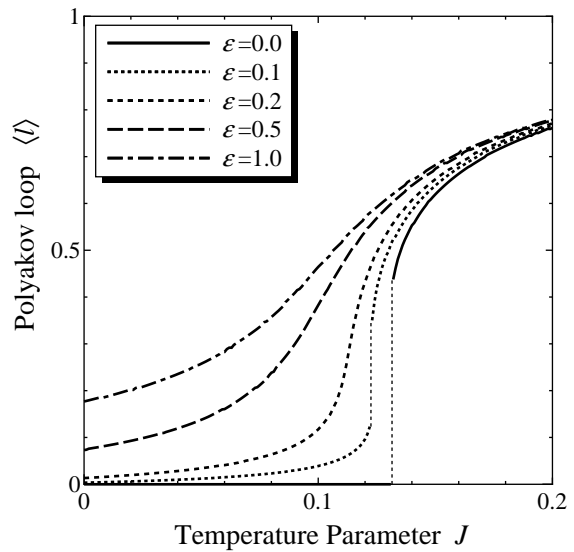}
 \caption{SU(3) (traced) Polyakov loop $\langle\ell\rangle$ as a
 function of the temperature parameter $J$ at various density
 parameters; $\epsilon=$0 (solid), 0.1 (dotted), 0.2 (short-dashed),
 0.5 (dashed), and 1.0 (dotted-dashed).  The first-order phase
 transition at $\epsilon=0$ and $\epsilon=0.1$ occurs at
 $J_{\text{c}}=0.132$ and $J_{\text{c}}=0.123$, respectively.}
 \label{fig:Jdep_su3}
\end{figure}

  Let us start with checking the monotonous correlation between
$\epsilon$ and $n$ as in the SU(2) case.  This allows us to regard
increase (or decrease) in the density parameter $\epsilon$ as increase
(or decrease) in the quark number density $n$.  The positive
correlation is obvious from the results we show in
Fig.~\ref{fig:edep_density_su3}.  The ``temperature'' or $J$
dependence is slightly greater than the SU(2) results.  We can give a
possible account for this as follows;  in the confined phase at small
$J$, the quark number density is suppressed as compared with high $J$
results.  This suppression comes from the group integration that
forces the thermally excited particles to be not quarks but (nearly)
color-singlet baryons consisting of $\Nc$
quarks~\cite{go_model,Oleszczuk:1992yg}.  In general larger $\Nc$
leads to stronger suppression by heavier excitation quanta.
Therefore, the stronger $J$ dependence presented in
Fig.~\ref{fig:edep_density_su3} originates from the stronger
suppression at small $J$.  The suppression is physically interpreted
as \textit{effective} tendency toward
confinement~\cite{Fukushima:2003fw}.

  Figure~\ref{fig:Jdep_su3} is the Polyakov loop as a function of $J$
corresponding to the SU(2) result in Fig.~\ref{fig:Jdep_su2} and to be
compared qualitatively with the lattice result of Fig.~7 in
Ref.~\cite{Blum:1995cb}.  We find a first-order phase transition for
$\epsilon=0$ at $J=J_{\text{c}}=0.132$ and for $\epsilon=0.1$ at
$J=J_{\text{c}}=0.123$.  The effect of nonzero $\epsilon$ smears the
transitional behavior and the phase transition eventually ceases to be
of first-order at a certain $\epsilon$.  The end-point of the
first-order phase boundary is a second-order critical point called the
critical end-point, which is of much interest in attempts to clarify
the QCD phase diagram~\cite{CEP,Asakawa:1989bq}.  We have crossover at
larger $\epsilon$.  The global picture is well consistent with what
has been already clarified in the Potts system as a toy model of
finite temperature and density QCD~\cite{Alford:2001ug,Kim:2005ck}.

\begin{figure}
 \includegraphics[width=7.5cm]{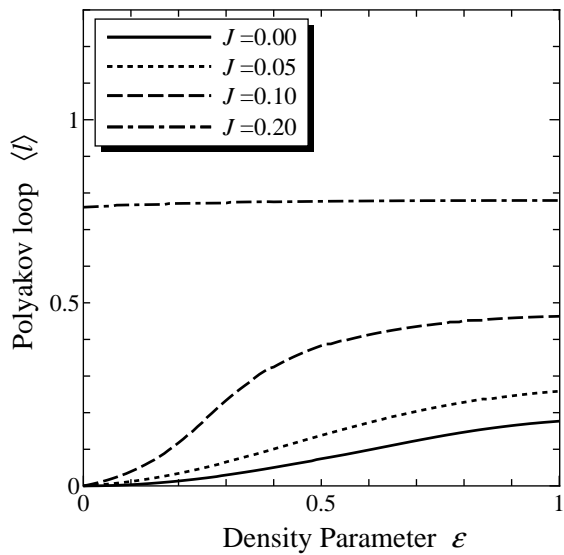}
 \caption{SU(3) Polyakov loop $\langle\ell\rangle$ as a function the
 density parameter $\epsilon$ at various temperature parameters; $J$=0
 (solid), 0.05 (dotted), 0.10 (short-dashed), 0.13 (dashed), and 0.2
 (dotted-dashed).  The gross feature is similar to the SU(2) case in
 Fig.~\ref{fig:edep_su2}.}
 \label{fig:edep_su3}
\end{figure}

  We plot the ``density'' dependence of the SU(3) Polyakov loop in
Fig.~\ref{fig:edep_su3} which is the SU(3) counterpart of
Fig.~\ref{fig:edep_su2}.  The SU(2) and SU(3) results are
qualitatively similar except for that the Polyakov loop is suppressed
at small $J$ and $\epsilon$ just as we found in the quark number
density.  It would be interesting if we could compare our results with
the lattice data, but unfortunately, the SU(3) data as a function of
$\epsilon$ is not available from Ref.~\cite{Blum:1995cb}.  We cannot
argue the $J$ (or $6/g^2$) dependence because it involves unknown
renormalization effects.

\begin{figure}
 \includegraphics[width=7.5cm]{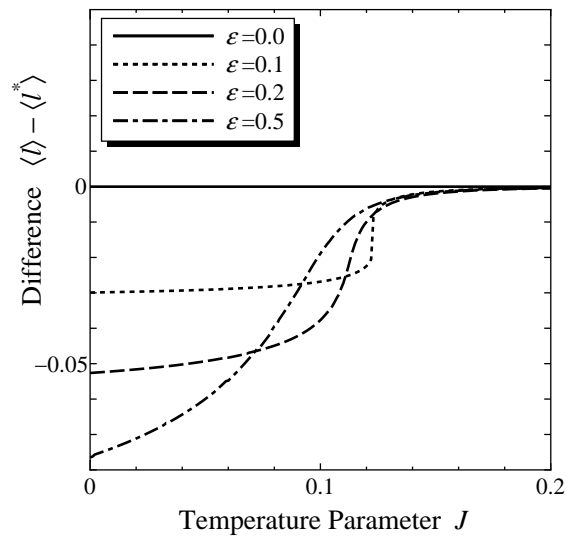}
 \caption{Difference of the SU(3) Polyakov loops $\langle\ell\rangle$
 and $\langle\ell^\ast\rangle$ as a function of the temperature
 parameter $J$ at various density parameters; $\epsilon=$0 (solid)
 which is zero entirely, 0.1 (dotted), 0.2 (short-dashed), and 0.5
 (dashed).}
 \label{fig:diff}
\end{figure}

\begin{figure}
 \includegraphics[width=7.8cm]{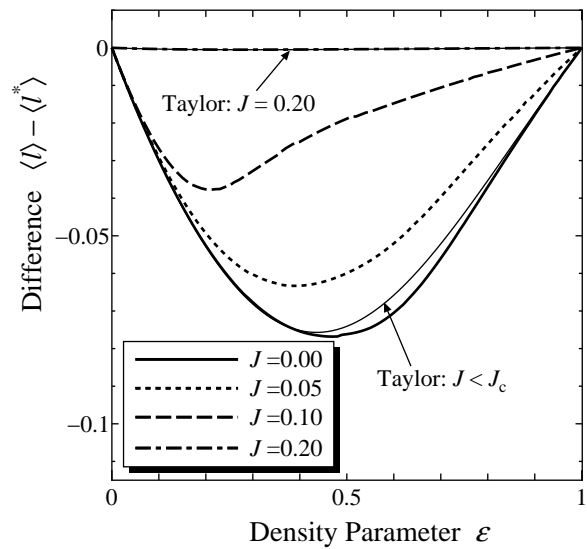}
 \caption{Difference of the SU(3) Polyakov loops $\langle\ell\rangle$
 and $\langle\ell^\ast\rangle$ as a function of the density parameter
 $\epsilon$ at various temperature parameters; $J=$0 (solid), 0.05
 (short-dashed), 0.1 (dashed), and 0.2 (dotted-dashed).  The thin
 curves represent the results from the Taylor expansion method in the
 confine phase at $J<J_{\text{c}}$ and in the deconfined phase at
 $J=0.20>J_{\text{c}}$ which is almost overlaid on the result from the
 phase reweighting.}
 \label{fig:edep_diff}
\end{figure}

\begin{figure*}
 \includegraphics[width=14cm]{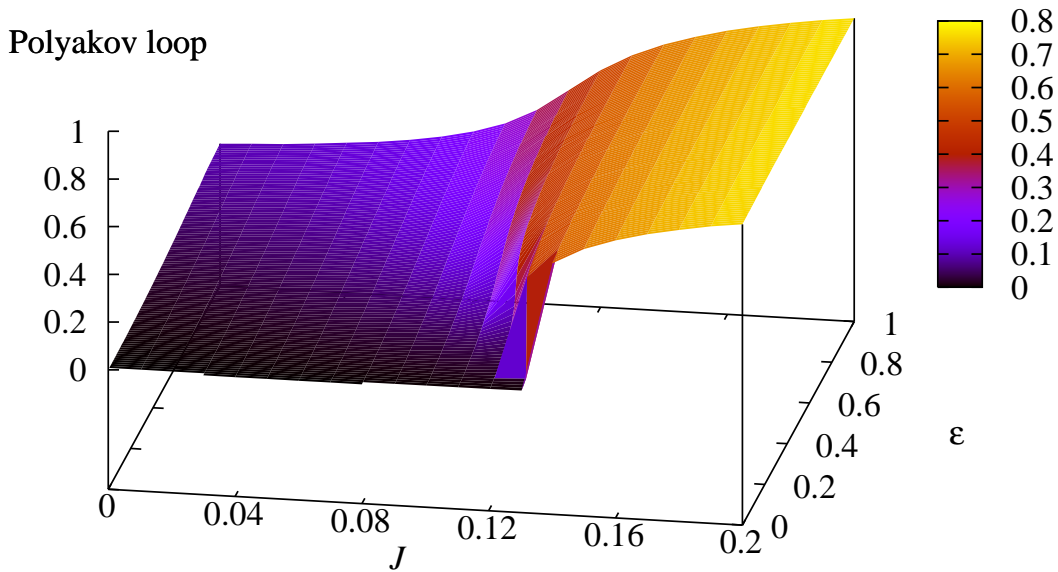}
 \caption{Three-dimensional plot of the fundamental Polyakov loop in
 the SU(3) case as a function of the temperature parameter $J$ and the
 density parameter $\epsilon$.}
 \label{fig:3d_su3}
\end{figure*}

  In the phase reweighting calculation we can see how
$\langle\ell\rangle$ and $\langle\ell^\ast\rangle$ become distinct at
$\mu_q\neq0$.  The observable $\ell-\ell^\ast$ is $C$-odd and so the
imaginary part of the fermion determinant is responsible for a
nonvanishing difference.  When $\epsilon$ is small in the fermionic
determinant (\ref{eq:sf_su3}) the imaginary part comes from
$\text{Im}\,\epsilon\ell\propto \epsilon(\ell-\ell^\ast)$.
Consequently the expectation value of the difference is proportional
to $\epsilon\langle(\text{Im}\,\ell)^2\rangle_0$ where
$\langle\cdots\rangle_0$ is taken at zero
density~\cite{Dumitru:2005ng}.  In Fig.~\ref{fig:diff} we present our
numerical results for the difference
$\langle\ell\rangle-\langle\ell^\ast\rangle$ as a function of $J$.
The difference is trivially zero at $\epsilon=0$ and $\epsilon=1$
where the fermion determinant is real.  As long as the density
parameter stays smaller than $\epsilon\sim0.5$, a larger density
parameter $\epsilon$ leads to a bigger difference.  For example, we
find the difference at $\epsilon=0.5$ as large as
$\langle\ell\rangle-\langle\ell^\ast\rangle=-0.076$ which is
comparable to $\langle\ell\rangle=0.073$.

  One can intuitively understand why $\langle\ell^\ast\rangle$ is
greater than $\langle\ell\rangle$ at nonzero $\mu_q$, which agrees
with what has been observed in the lattice simulation~\cite{Taylor}.
It is because, as discussed also in Ref.~\cite{Taylor}, the presence
of quarks at finite density enhances the screening effect for
antiquarks so that the antiquark excitation costs less energy.

  The dense-heavy model, or $e^{-\Sf}$, originally takes a form of
power series in $\epsilon$ as seen in the
expression~(\ref{eq:sf_su3}).  We have thus performed the Taylor
expansion also to estimate the Polyakov loop difference.  The
expectation value of the coefficient of the $\epsilon$ series is
calculated at zero density $\epsilon=0$.  In this static model the
fermionic contribution is just vanishing at $\epsilon=0$, and so the
Taylor expansion method gives rise to the identical output for any
$J<J_{\text{c}}$ in the confined phase where $x$ remains zero.  In the
deconfined phase $J>J_{\text{c}}$ the $J$ dependence is brought in by
nonzero $x$.  Figure~\ref{fig:edep_diff} shows the difference
$\langle\ell\rangle-\langle\ell^\ast\rangle$ as a function of
$\epsilon$ at various $J$ with the Taylor expansion results in the
confined and deconfined phase.  We can see excellent agreement between
two methods from the comparison at $J=0.00$ and $J=0.20$.  However,
the Taylor expansion cannot reproduce the results at
$0<J<J_{\text{c}}$ at all.  The lesson from our model study is that
the Taylor expansion in terms of \textit{density} breaks down in the
presence of the first-order phase transition with respect to
\textit{temperature}.  Still, in the deconfined phase at high
temperature, the expansion is validated.  We do not think that this
finding is trivial in our model.  In the realistic QCD lattice
simulation with $2+1$ flavors, the zero density result is most likely
crossover, and thus the Taylor expansion method should be reliable at
all temperatures.

  As we showed in the previous subsection, we shall present the
three-dimensional plot of the SU(3) Polyakov loop as a function of the
temperature parameter $J$ and the density parameter $\epsilon$ in
Fig.~\ref{fig:3d_su3}.  The figure is qualitatively consistent with
the lattice result shown in Fig.~5 in Ref.~\cite{Blum:1995cb}.

\begin{figure}
 \includegraphics[width=7.5cm]{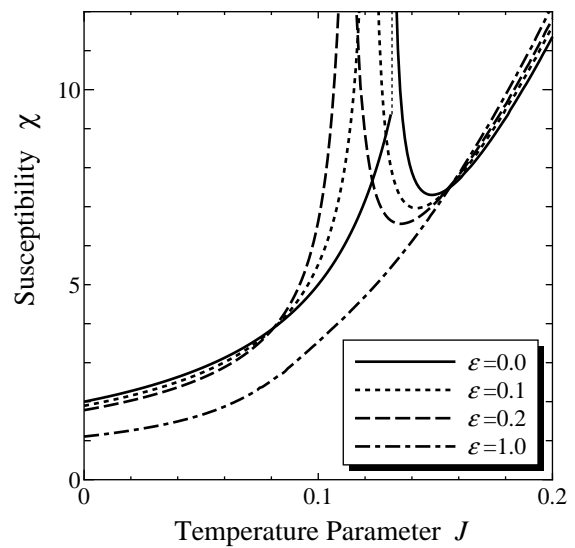}
 \caption{Susceptibility relevant to the SU(3) Polyakov loop $\ell$ as
 a function of $J$ at various density parameters; $\epsilon=0.0$
 (solid), 0.1 (short-dashed), 0.2 (dashed), and 1.0 (dotted-dashed).}
 \label{fig:Jdep_sus_su3}
\end{figure}

\begin{figure}
 \includegraphics[width=7.5cm]{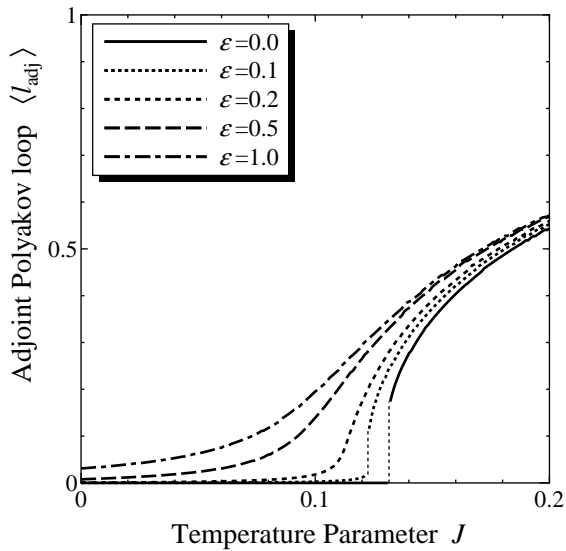}
 \caption{SU(3) adjoint Polyakov loop
 $\langle\ell_{\text{adj}}\rangle$ as a function of the temperature
 parameter $J$ at various density parameters; $\epsilon$=0.0 (solid),
 0.1 (dotted), 0.2 (short-dashed), 0.5 (dashed), and 1.0
 (dotted-dashed).}
 \label{fig:Jdep_ad_su3}
\end{figure}

  We shall turn to the susceptibility $\chi$ to examine the phase
transition more closely.  Because the phase transition at small
$\epsilon$ is of first-order, $\chi$ jumps discontinuously at
$J=J_{\text{c}}$ as seen in Fig.~\ref{fig:Jdep_sus_su3}.  The
susceptibility $\chi$ grows as the model parameters approach the
critical end-point where the second-order phase transition takes
place.  Figure~\ref{fig:Jdep_sus_su3} implies that the zero density
($\epsilon=0$) result is significantly affected by the critical
end-point which should be located nearby at small $\epsilon$.  In view
of Figs.~\ref{fig:Jdep_sus_su2} and \ref{fig:Jdep_sus_su3} the
critical region in the SU(3) case is wider than the SU(2) case.

  Let us comment on the adjoint Polyakov loop here again.  As we
mentioned in discussions on the SU(2) results, the adjoint Polyakov
loop should require proper renormalization beyond the mean-field
treatment.  We show the adjoint Polyakov loop result in
Fig.~\ref{fig:Jdep_ad_su3} just because we can do that.  The
first-order phase transition is located at the same point;
$J=J_{\text{c}}\simeq0.132$, of course.  The group integration over
$L$ suppresses $\langle\ell_{\text{adj}}\rangle$ at low $J$ even for
large $\epsilon$, which makes the crossover in
Fig.~\ref{fig:Jdep_ad_su3} look shaper than in case of the fundamental
Polyakov loop in Fig.~\ref{fig:Jdep_su3}.

\begin{figure}
 \includegraphics[width=7.5cm]{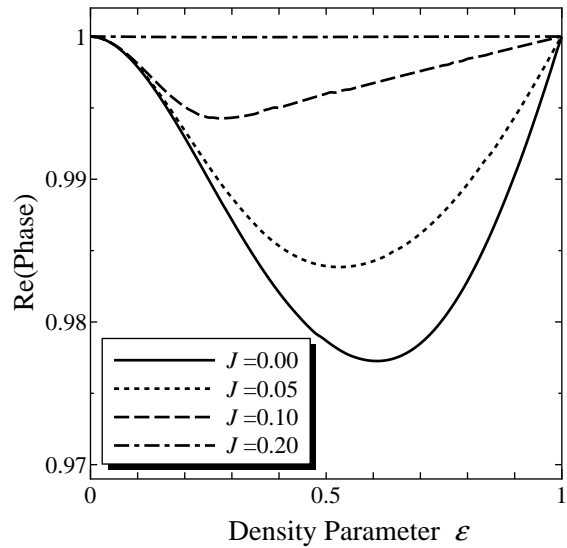}
 \caption{Phase of the fermion determinant
  $\text{Re}\langle e^{-i\theta}\rangle$ per lattice site as a
 function of the density parameter $\epsilon$ at various temperature
 parameters; $J=$0 (solid), 0.05 (short-dashed), 0.10 (dashed), and
 0.20 (dotted-dashed).}
 \label{fig:edep_phase}
\end{figure}

  Finally we present the results for the expectation value of the
phase factor of the fermionic determinant, $e^{-i\Theta}$.  We plot
$\text{Re}\langle e^{-i\theta}\rangle$ as a function of $\epsilon$ in
Fig.~\ref{fig:edep_phase}, where $\theta$ is the phase at each lattice
site;
$\theta\equiv-\arg(1+\epsilon^3+3\epsilon\ell+3\epsilon^2\ell^\ast)$
(i.e.\ $\Theta=\sum_{\vec{x}}\theta$.)
Comparing it with Fig.~9 in Ref.~\cite{Blum:1995cb}, we can surely
check that our results qualitatively reproduce the lattice data.  For
more quantitative arguments, let us put the volume $6^3=216$ of the
lattice simulation in Ref.~\cite{Blum:1995cb} into our results.  The
expected phase factor $\langle e^{-i\Theta}\rangle$ can be, as a rough
estimate, approximated as $(\langle e^{-i\theta}\rangle)^{216}$.  For
instance, our $J=0$ result has a minimum at $\epsilon=0.61$ where
$\langle e^{-i\theta}\rangle\simeq 0.977$, and we get
$0.977^{216}=0.0066$.  The minimum value in Fig.~9 in
Ref.~\cite{Blum:1995cb} is read as of order $0.01$, which is not quite
far from our estimate viewed on the logarithmic plot.


\subsection{Mean-Field Free Energy}
\label{sec:MFFreeEnergy}

  Here we will pursue another strategy to deal with the difference
between $\langle\ell\rangle$ and $\langle\ell^\ast\rangle$ as a
double-check.  It is possible to extend the mean-field ansatz as
\begin{equation}
 \Smf[L] = -\frac{x}{2}\sum_{\vec{x}}\bigl[\ell(\vec{x})
  \!+\!\ell^\ast(\vec{x})\bigr] -\frac{y}{2}\sum_{\vec{x}}\bigl[
  \ell(\vec{x})\!-\!\ell^\ast(\vec{x})\bigr] \,,
\label{eq:ansatz2}
\end{equation}
and then we can compute the mean-field free energy from
Eq.~(\ref{eq:free_energy}) as a function of $x$ and $y$.  After the
integration over $L$, the free energy $\fmf(x,y)$ is a real function
of real variables $x$ and $y$.  We can thus fix the mean-fields by
\begin{equation}
 \frac{\partial\fmf}{\partial x}\biggr|_{(x,y)=(x_0,y_0)}
 =\frac{\partial\fmf}{\partial y}\biggr|_{(x,y)=(x_0,y_0)}=0 \,.
\label{eq:mean_field_eq}
\end{equation}
Nonzero $y_0$ appears in the presence of nonzero $\mu_q$.  It turns
out that around $(x_0,y_0)$ the free energy $\fmf(x,y)$ has a minimum
in the $x$ direction, while it has a maximum in the $y$ direction.
That is, the solution to Eq.~(\ref{eq:mean_field_eq}) is a
saddle-point of $\fmf(x,y)$, which is consistent with
Ref.~\cite{Dumitru:2005ng}.  The instability with respect to $y$
should be a remnant of the sign problem~\footnote{One way to
understand instability in $y$ might be that $\ell-\ell^\ast$ is pure
imaginary though $y$ corresponding to
$\langle\ell\rangle-\langle\ell^\ast\rangle$ is real.}  We here point
out that this instability also exists in the zero density limit;
$\fmf(x,y)$ has a saddle-point at $y=0$ even at zero density.  It
should be instructive to take a closer look at how the saddle-point
arises even at zero density.  If we expand the free energy in terms of
$\langle\ell\rangle$ and $\langle\ell^\ast\rangle$, the leading term
dependent on $\langle\ell\rangle$ and $\langle\ell^\ast\rangle$ is
quadratic $\sim\langle\ell\rangle\langle\ell^\ast\rangle$ as explicitly
seen in a simple estimate in Ref.~\cite{Ratti:2005jh}.  This form of
the free energy implies an instability inducing
$\langle\ell\rangle\neq\langle\ell^\ast\rangle$ because it can written
as $\langle\ell\rangle\langle\ell^\ast\rangle \propto
(\langle\ell\rangle\!+\!\langle\ell^\ast\rangle)^2
-(\langle\ell\rangle\!-\!\langle\ell^\ast\rangle)^2$.

  The situation is somewhat analogous to thermodynamics in the finite
density NJL-model calculation.  If the free energy $f$ is calculated
as a function of the renormalized chemical potential
$\mu_q$~\cite{Asakawa:1989bq}, then the value of $\mu_q$ is fixed by
the condition $\partial f/\partial\mu_q=0$ which corresponds to not a
minimum but a maximum of $f$ as a function of $\mu_q$.  This is not
problematic, however, because the condition $\partial
f/\partial\mu_q=0$ means a \textit{constraint} equation of number
density.  Likewise, we might as well think that the determination of
$y$ in the Polyakov loop dynamics is not energetic but the second
equation of (\ref{eq:mean_field_eq}) has something to do with a
constraint equation of number density of gauge charge, namely, the
Gauss law constraint.  If this conjecture is the case, though the
rigorous proof is beyond our current scope, the saddle-point nature of
the free energy is no longer an obstacle to fix $x$ and $y$.

\begin{figure}
 \includegraphics[width=7.5cm]{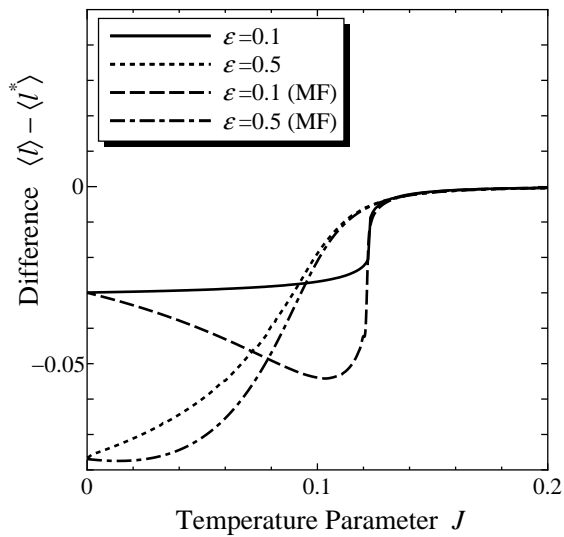}
 \caption{Comparison to the difference
 $\langle\ell\rangle-\langle\ell^\ast\rangle$ estimated by the
 saddle-point of the mean-field free energy at various densities;
 $\epsilon=$0.1 (solid and dashed) and 0.5 (dotted and
 dotted-dashed).}
 \label{fig:diff_comp}
\end{figure}

  Let us see what comes out if we calculate the Polyakov loop using
the ansatz (\ref{eq:ansatz2}) with the solution to
Eq.~(\ref{eq:mean_field_eq}) assuming that the second equation of
(\ref{eq:mean_field_eq}) stems from a constraint associated with gauge
dynamics.  In Fig.~\ref{fig:diff_comp} we show the difference
$\langle\ell\rangle-\langle\ell^\ast\rangle$ as a function of $J$.
The solid and dotted curves at $\epsilon=0.1$ and $\epsilon=0.5$
respectively are just the same as already presented in
Fig.~\ref{fig:diff}.  The dashed and dotted-dashed curves are the
counterparts derived from the mean-field energy with $x$ and $y$.

  In view of two $\epsilon=0.5$ curves, on the one hand, the
mean-field result is entirely consistent with what we found in the
phase reweighting method.  On the other hand, the coincidence is not
very good for the $\epsilon=0.1$ results except for $J\simeq0$ and
$J>J_{\text{c}}$.  This discrepancy comes from a singularity of the
dense-heavy model located at $\epsilon=0$;  when $\epsilon$ is small,
the fermion action in the dense-heavy model is approximated as
$\Sf[L]\sim-3\epsilon\sum_{\vec{x}}\ell(\vec{x})$.  Hence, nonzero
$\epsilon$ tends to align the vacuum into the direction of $x=y$ (see
Eq.~(\ref{eq:ansatz2})), and the model does not reduce to a pure
gluonic theory with $y=0$ smoothly in the limit of $\epsilon\to0^+$.
This situation makes a sharp contrast to finite density QCD.  In the
strong coupling expansion, for a simple example, the fermion action is
$\Sf[L]\sim H\sum_{\vec{x}}\bigl[\ell(\vec{x})e^{\mu_q/T}+\ell^\ast
(\vec{x})e^{-\mu_q/T}\bigr]$.  In the limit of $\mu_q\to0$,
therefore, the fermionic action acts as an external field toward
$x\neq0$ and $y=0$, so that the vacuum alignment of a pure gluonic
theory is smoothly retrieved in the $\mu_q\to0$ and $H\to0^+$ limit.

  Apart from the discrepancy inherent to the singular behavior of the
dense-heavy model near $\epsilon=0$, the mean-field free energy leads
to the results in accord with the phase reweighting method.  This is
an indirect evidence for that the saddle-point is not harmful
actually, as we conjectured.

  We shall comment on a possible clue to resolve the sign problem
based on what we have seen here.  We changed the sign problem into a
form of the saddle-point of the mean-field free energy.  The
saddle-point appears harmless from our analyses, presumably stabilized
by the Gauss law, and then the sign problem is resolved.  Of course,
it is highly nontrivial how to map this analytical procedure to the
lattice QCD simulation.  Still, we would point out that the clustering
method developed in Ref.~\cite{Alford:2001ug} is a philosophically
similar idea along this line;  partial integration over the cluster
domains wipes away the sign problem, just like seen in our mean-field
free energy free from the sign problem after the group integration.

  The field-theoretical approach to reveal the relation between the
saddle-point of the free energy and the Gauss law is beyond our
current scope, but it definitely deserves further investigation.


\section{Summary}
\label{sec:summary}

  We investigated the dense-heavy model and observed the sign problem
at finite density within the framework of the mean-field
approximation.  We calculated the quark number density, the
fundamental and adjoint Polyakov loop, the susceptibility, and the
phase of the fermion determinant as a function of the model parameters
specifying the temperature and the density.  All the mean-field
results are reasonable and even in quantitative agreement with the
lattice data.

  In the environment free from the sign problem in the SU(2) case we
found that the mean-field approximation goes better than expected.
Our mean-field results nicely reproduced the quark number density and
the Polyakov loop measured on the lattice once the unknown Polyakov
loop renormalization is fixed by fitting.  Then, we proceeded into the
SU(3) case where the sign problem is relevant.

  We saw that the approximation scheme which is capable of describing
the different $\mu_q$ dependence of $\langle\ell\rangle$ and
$\langle\ell^\ast\rangle$ cannot avoid the sign problem even at the
mean-field level.  We applied the phase reweighting method as a
practical resolution in order to handle the complex fermion
determinant.  We acquired $\langle\ell\rangle-\langle\ell^\ast\rangle$
as a function of the density parameter or the temperature parameter.
So far, there are not many lattice data available for this quantity,
but our results $\langle\ell^\ast\rangle>\langle\ell\rangle$ are
consistent with other model studies as well as what has been obtained
from the Taylor expansion method on the lattice.

  The important message from our work is the following.  The sign
problem may be a serious obstacle even in the mean-field model studies
at finite temperature and density.  We would say, at least, that one
should be careful enough when one deals with the Polyakov loop
behavior at finite density.  The chiral effective models with the
Polyakov loop coupled are examples that definitely need more or less
caution for application to the finite density problem.  The phase
reweighting method is one prescription in order to approximate the
expectation value in a manageable way.  This prescription is, however,
only practical and not any solution to the sign problem.  We would
insist that one cannot solve the QCD sign problem until one can at
least find a solution to the sign problem appearing even in that
simple model setting.  The converse is not necessarily true, though.

  In the future, we would like to apply other ideas than the phase
reweighting method for the sign problem at the mean-field level.  The
bottom line here is, thus, that we have formulated an analytical
testing ground to think about the sign problem, and we believe that
our clarification would be useful for further developments.


\begin{acknowledgments}
The authors thank Y.~Aoki, T.~Blum, Ph.~de~Forcrand, R.~Pisarski,
C.~Schmidt for useful conversations.  The authors specially thank
S.~Ejiri for discussions on the dense-heavy model which urged us to
initiate this work.  This research was supported in part by RIKEN BNL
Research Center and the U.S.\ Department of Energy under cooperative
research agreement \#DE-AC02-98CH10886.
\end{acknowledgments}



\begin{thebibliography}{99}

\bibitem{review}
For theoretical reviews on QCD in extreme environments, see,
  H.~Meyer-Ortmanns,
  Rev.\ Mod.\ Phys.\  {\bf 68}, 473 (1996)
  [arXiv:hep-lat/9608098];
  F.~Wilczek,
  arXiv:hep-ph/0003183;
  D.~H.~Rischke,
  Prog.\ Part.\ Nucl.\ Phys.\  {\bf 52}, 197 (2004)
  [arXiv:nucl-th/0305030].

\bibitem{sQGP}
For reviews on recent experimental data and interpretation, see,
  M.~Gyulassy and L.~McLerran,
  Nucl.\ Phys.\ A {\bf 750}, 30 (2005)
  [arXiv:nucl-th/0405013];
  J.~L.~Nagle,
  arXiv:nucl-th/0608070.

\bibitem{CSC}
For reviews on color superconductivity and the phase diagram of cold
and dense quark matter, see,
  K.~Rajagopal and F.~Wilczek,
  arXiv:hep-ph/0011333;
  K.~Fukushima,
  arXiv:hep-ph/0510299;
  M.~G.~Alford,
  arXiv:hep-lat/0610046;
  S.~B.~Ruster, V.~Werth, M.~Buballa, I.~A.~Shovkovy and D.~H.~Rischke,
  arXiv:nucl-th/0602018.

\bibitem{lattice}
For reviews on lattice QCD simulations, see,
  F.~Karsch,
  Lect.\ Notes Phys.\  {\bf 583}, 209 (2002)
  [arXiv:hep-lat/0106019];
  K.~Kanaya,
  Nucl.\ Phys.\ A {\bf 715}, 233 (2003)
  [arXiv:hep-ph/0209116];
  E.~Laermann and O.~Philipsen,
  Ann.\ Rev.\ Nucl.\ Part.\ Sci.\  {\bf 53}, 163 (2003)
  [arXiv:hep-ph/0303042];
  C.~Schmidt and T.~Umeda,
  arXiv:hep-lat/0609032.

\bibitem{Tc}
  M.~Cheng {\it et al.},
  Phys.\ Rev.\ D {\bf 74}, 054507 (2006)
  [arXiv:hep-lat/0608013];
  Y.~Aoki, Z.~Fodor, S.~D.~Katz and K.~K.~Szabo,
  Phys.\ Lett.\ B {\bf 643}, 46 (2006)
  [arXiv:hep-lat/0609068].

\bibitem{Aoki:2005vt}
  Y.~Aoki, Z.~Fodor, S.~D.~Katz and K.~K.~Szabo,
  JHEP {\bf 0601}, 089 (2006)
  [arXiv:hep-lat/0510084].

\bibitem{Bernard:2004je}
  C.~Bernard {\it et al.}  [MILC Collaboration],
  Phys.\ Rev.\ D {\bf 71}, 034504 (2005)
  [arXiv:hep-lat/0405029].

\bibitem{charmonium}
  T.~Umeda, K.~Nomura and H.~Matsufuru,
  Eur.\ Phys.\ J.\ C {\bf 39S1}, 9 (2005)
  [arXiv:hep-lat/0211003];
  M.~Asakawa and T.~Hatsuda,
  Phys.\ Rev.\ Lett.\  {\bf 92}, 012001 (2004)
  [arXiv:hep-lat/0308034];
  S.~Datta, F.~Karsch, P.~Petreczky and I.~Wetzorke,
  Phys.\ Rev.\ D {\bf 69}, 094507 (2004)
  [arXiv:hep-lat/0312037].

\bibitem{review:signproblem}
For reviews on the sign problem, see,
  S.~Muroya, A.~Nakamura, C.~Nonaka and T.~Takaishi,
  Prog.\ Theor.\ Phys.\  {\bf 110}, 615 (2003)
  [arXiv:hep-lat/0306031];
  M.~P.~Lombardo,
  Prog.\ Theor.\ Phys.\ Suppl.\  {\bf 153}, 26 (2004)
  [arXiv:hep-lat/0401021];
  S.~Aoki,
  Int.\ J.\ Mod.\ Phys.\ A {\bf 21}, 682 (2006)
  [arXiv:hep-lat/0509068].

\bibitem{reweighting}
  Z.~Fodor and S.~D.~Katz,
  Phys.\ Lett.\ B {\bf 534}, 87 (2002)
  [arXiv:hep-lat/0104001];
  JHEP {\bf 0203}, 014 (2002)
  [arXiv:hep-lat/0106002];
  JHEP {\bf 0404}, 050 (2004)
  [arXiv:hep-lat/0402006].

\bibitem{Taylor}
  C.~R.~Allton {\it et al.},
  Phys.\ Rev.\ D {\bf 66}, 074507 (2002)
  [arXiv:hep-lat/0204010];
  C.~R.~Allton, S.~Ejiri, S.~J.~Hands, O.~Kaczmarek, F.~Karsch, E.~Laermann and C.~Schmidt,
  Phys.\ Rev.\ D {\bf 68}, 014507 (2003)
  [arXiv:hep-lat/0305007].

\bibitem{imaginary}
  M.~G.~Alford, A.~Kapustin and F.~Wilczek,
  Phys.\ Rev.\ D {\bf 59}, 054502 (1999)
  [arXiv:hep-lat/9807039];
  P.~de Forcrand and O.~Philipsen,
  Nucl.\ Phys.\ B {\bf 642}, 290 (2002)
  [arXiv:hep-lat/0205016];
  Nucl.\ Phys.\ B {\bf 673}, 170 (2003)
  [arXiv:hep-lat/0307020];
  [arXiv:hep-lat/0607017];
  M.~D'Elia and M.~P.~Lombardo,
  Phys.\ Rev.\ D {\bf 67}, 014505 (2003)
  [arXiv:hep-lat/0209146].

\bibitem{polyakov_loop}
  A.~M.~Polyakov,
  Phys.\ Lett.\ B {\bf 72}, 477 (1978);
  L.~Susskind,
  Phys.\ Rev.\ D {\bf 20}, 2610 (1979);
  B.~Svetitsky and L.~G.~Yaffe,
  Nucl.\ Phys.\ B {\bf 210}, 423 (1982);
  B.~Svetitsky,
  Phys.\ Rept.\  {\bf 132}, 1 (1986).

\bibitem{polyakov_lattice}
  L.~D.~McLerran and B.~Svetitsky,
  Phys.\ Lett.\ B {\bf 98}, 195 (1981);
  L.~D.~McLerran and B.~Svetitsky,
  Phys.\ Rev.\ D {\bf 24}, 450 (1981);
  J.~Kuti, J.~Polonyi and K.~Szlachanyi,
  Phys.\ Lett.\ B {\bf 98}, 199 (1981).

\bibitem{weiss}
  N.~Weiss,
  Phys.\ Rev.\ D {\bf 24}, 475 (1981);
  Phys.\ Rev.\ D {\bf 25}, 2667 (1982);
  V.~M.~Belyaev,
  Phys.\ Lett.\ B {\bf 241}, 91 (1990);
  Phys.\ Lett.\ B {\bf 254}, 153 (1991);
  C.~P.~Korthals Altes,
  Nucl.\ Phys.\ B {\bf 420}, 637 (1994)
  [arXiv:hep-th/9310195];
  K.~Fukushima and K.~Ohta,
  J.\ Phys.\ G {\bf 26}, 1397 (2000)
  [arXiv:hep-ph/0011108];

\bibitem{strong}
  J.~Polonyi and K.~Szlachanyi,
  Phys.\ Lett.\ B {\bf 110}, 395 (1982);
  M.~Gross, J.~Bartholomew and D.~Hochberg,
  ``SU(N) Deconfinement Transition And The N State Clock Model,''
  EFI-83-35-CHICAGO.

\bibitem{pl_model}
  A.~Dumitru and R.~D.~Pisarski,
  Phys.\ Lett.\ B {\bf 504}, 282 (2001)
  [arXiv:hep-ph/0010083];
  Phys.\ Lett.\ B {\bf 525}, 95 (2002)
  [arXiv:hep-ph/0106176];
  Phys.\ Rev.\ D {\bf 66}, 096003 (2002)
  [arXiv:hep-ph/0204223].

\bibitem{go_model}
  A.~Gocksch and M.~Ogilvie,
  Phys.\ Rev.\ D {\bf 31}, 877 (1985);
  K.~Fukushima,
  Phys.\ Lett.\ B {\bf 553}, 38 (2003)
  [arXiv:hep-ph/0209311];
  Phys.\ Rev.\ D {\bf 68}, 045004 (2003)
  [arXiv:hep-ph/0303225].

\bibitem{hatta}
  Y.~Hatta and K.~Fukushima,
  Phys.\ Rev.\ D {\bf 69}, 097502 (2004)
  [arXiv:hep-ph/0307068];
  A.~Mocsy, F.~Sannino and K.~Tuominen,
  Phys.\ Rev.\ Lett.\  {\bf 92}, 182302 (2004)
  [arXiv:hep-ph/0308135].

\bibitem{Fukushima:2003fw}
  K.~Fukushima,
  Phys.\ Lett.\ B {\bf 591}, 277 (2004)
  [arXiv:hep-ph/0310121].

\bibitem{KorthalsAltes:1999cp}
  C.~P.~Korthals Altes, R.~D.~Pisarski and A.~Sinkovics,
  Phys.\ Rev.\ D {\bf 61}, 056007 (2000)
  [arXiv:hep-ph/9904305].

\bibitem{Weiss:1987mp}
  N.~Weiss,
  Phys.\ Rev.\ D {\bf 35}, 2495 (1987).

\bibitem{fnjl}
  E.~Megias, E.~Ruiz Arriola and L.~L.~Salcedo,
  Phys.\ Rev.\ D {\bf 74}, 065005 (2006)
  [arXiv:hep-ph/0412308];
  S.~K.~Ghosh, T.~K.~Mukherjee, M.~G.~Mustafa and R.~Ray,
  Phys.\ Rev.\ D {\bf 73}, 114007 (2006)
  [arXiv:hep-ph/0603050];
  H.~Hansen, W.~M.~Alberico, A.~Beraudo, A.~Molinari, M.~Nardi and C.~Ratti,
  arXiv:hep-ph/0609116;
  S.~Mukherjee, M.~G.~Mustafa and R.~Ray,
  arXiv:hep-ph/0609249.

\bibitem{Ratti:2005jh}
  C.~Ratti, M.~A.~Thaler and W.~Weise,
  Phys.\ Rev.\ D {\bf 73}, 014019 (2006)
  [arXiv:hep-ph/0506234];
  S.~Roessner, C.~Ratti and W.~Weise,
  arXiv:hep-ph/0609281.

\bibitem{Dumitru:2005ng}
  A.~Dumitru, R.~D.~Pisarski and D.~Zschiesche,
  Phys.\ Rev.\ D {\bf 72}, 065008 (2005)
  [arXiv:hep-ph/0505256].

\bibitem{Blum:1995cb}
  T.~C.~Blum, J.~E.~Hetrick and D.~Toussaint,
  Phys.\ Rev.\ Lett.\  {\bf 76}, 1019 (1996)
  [arXiv:hep-lat/9509002].
  (We note that the published version of this paper does not contain
  the SU(2) results which are present in the preprint version.  We
  refer only to the preprint version.)

\bibitem{Bender:1991gn}
  I.~Bender {\it et al.},
  Nucl.\ Phys.\ Proc.\ Suppl.\  {\bf 26}, 323 (1992);
  J.~Engels, O.~Kaczmarek, F.~Karsch and E.~Laermann,
  Nucl.\ Phys.\ B {\bf 558}, 307 (1999)
  [arXiv:hep-lat/9903030].

\bibitem{Son:2000xc}
  D.~T.~Son and M.~A.~Stephanov,
  Phys.\ Rev.\ Lett.\  {\bf 86}, 592 (2001)
  [arXiv:hep-ph/0005225].

\bibitem{su2}
  E.~Dagotto, F.~Karsch and A.~Moreo,
  Phys.\ Lett.\ B {\bf 169}, 421 (1986);
  J.~U.~Klatke and K.~H.~Mutter,
  Nucl.\ Phys.\ B {\bf 342}, 764 (1990);
  S.~Hands, J.~B.~Kogut, M.~P.~Lombardo and S.~E.~Morrison,
  Nucl.\ Phys.\ B {\bf 558}, 327 (1999)
  [arXiv:hep-lat/9902034];
  J.~B.~Kogut, D.~Toublan and D.~K.~Sinclair,
  Phys.\ Lett.\ B {\bf 514}, 77 (2001)
  [arXiv:hep-lat/0104010];
  J.~B.~Kogut, D.~Toublan and D.~K.~Sinclair,
  Nucl.\ Phys.\ B {\bf 642}, 181 (2002)
  [arXiv:hep-lat/0205019];
  J.~B.~Kogut, D.~Toublan and D.~K.~Sinclair,
  Phys.\ Rev.\ D {\bf 68}, 054507 (2003)
  [arXiv:hep-lat/0305003];
  S.~Muroya, A.~Nakamura and C.~Nonaka,
  Phys.\ Lett.\ B {\bf 551}, 305 (2003)
  [arXiv:hep-lat/0211010];
  B.~Vanderheyden and A.~D.~Jackson,
  Phys.\ Rev.\ D {\bf 64}, 074016 (2001)
  [arXiv:hep-ph/0102064];
  J.~B.~Kogut, M.~A.~Stephanov and D.~Toublan,
  Phys.\ Lett.\ B {\bf 464}, 183 (1999)
  [arXiv:hep-ph/9906346];
  J.~B.~Kogut, M.~A.~Stephanov, D.~Toublan, J.~J.~M.~Verbaarschot and A.~Zhitnitsky,
  Nucl.\ Phys.\ B {\bf 582}, 477 (2000)
  [arXiv:hep-ph/0001171];
  Y.~Nishida, K.~Fukushima and T.~Hatsuda,
  Phys.\ Rept.\  {\bf 398}, 281 (2004)
  [arXiv:hep-ph/0306066];
  S.~Chandrasekharan and F.~J.~Jiang,
  Phys.\ Rev.\ D {\bf 74}, 014506 (2006)
  [arXiv:hep-lat/0602031];
  B.~Alles, M.~D'Elia and M.~P.~Lombardo,
  Nucl.\ Phys.\ B {\bf 752}, 124 (2006)
  [arXiv:hep-lat/0602022];
  S.~Hands, S.~Kim and J.~I.~Skullerud,
  arXiv:hep-lat/0604004.

\bibitem{Hasenfratz:1984em}
  P.~Hasenfratz and F.~Karsch,
  Phys.\ Rept.\  {\bf 103}, 219 (1984).

\bibitem{Kogut:1981ez}
  J.~B.~Kogut, M.~Snow and M.~Stone,
  Nucl.\ Phys.\ B {\bf 200}, 211 (1982).

\bibitem{Engels:1989fz}
  J.~Engels, J.~Fingberg and M.~Weber,
  Nucl.\ Phys.\ B {\bf 332}, 737 (1990).

\bibitem{Dumitru:2003hp}
  A.~Dumitru, Y.~Hatta, J.~Lenaghan, K.~Orginos and R.~D.~Pisarski,
  Phys.\ Rev.\ D {\bf 70}, 034511 (2004)
  [arXiv:hep-th/0311223].

\bibitem{Lenz:1998qk}
  F.~Lenz and M.~Thies,
  Annals Phys.\  {\bf 268}, 308 (1998)
  [arXiv:hep-th/9802066].

\bibitem{deForcrand:1999cy}
  P.~de Forcrand and V.~Laliena,
  Phys.\ Rev.\ D {\bf 61}, 034502 (2000)
  [arXiv:hep-lat/9907004].

\bibitem{Oleszczuk:1992yg}
  M.~Oleszczuk and J.~Polonyi,
  Annals Phys.\  {\bf 227}, 76 (1993).

\bibitem{CEP}
  A.~Barducci, R.~Casalbuoni, S.~De Curtis, R.~Gatto and G.~Pettini,
  Phys.\ Lett.\ B {\bf 231}, 463 (1989);
  F.~Wilczek,
  Int.\ J.\ Mod.\ Phys.\ A {\bf 7}, 3911 (1992)
  [Erratum-ibid.\ A {\bf 7}, 6951 (1992)];
  A.~Barducci, R.~Casalbuoni, G.~Pettini and R.~Gatto,
  Phys.\ Rev.\ D {\bf 49}, 426 (1994);
  J.~Berges and K.~Rajagopal,
  Nucl.\ Phys.\ B {\bf 538}, 215 (1999)
  [arXiv:hep-ph/9804233];
  M.~A.~Halasz, A.~D.~Jackson, R.~E.~Shrock, M.~A.~Stephanov and J.~J.~M.~Verbaarschot,
  Phys.\ Rev.\ D {\bf 58}, 096007 (1998)
  [arXiv:hep-ph/9804290];
  M.~A.~Stephanov, K.~Rajagopal and E.~V.~Shuryak,
  Phys.\ Rev.\ Lett.\  {\bf 81}, 4816 (1998)
  [arXiv:hep-ph/9806219];
  M.~A.~Stephanov, K.~Rajagopal and E.~V.~Shuryak,
  Phys.\ Rev.\ D {\bf 60}, 114028 (1999)
  [arXiv:hep-ph/9903292];
  K.~Fukushima,
  Phys.\ Rev.\ C {\bf 67}, 025203 (2003)
  [arXiv:hep-ph/0209270];
  Y.~Hatta and T.~Ikeda,
  Phys.\ Rev.\ D {\bf 67}, 014028 (2003)
  [arXiv:hep-ph/0210284].

\bibitem{Asakawa:1989bq}
  M.~Asakawa and K.~Yazaki,
  Nucl.\ Phys.\ A {\bf 504}, 668 (1989);

\bibitem{Alford:2001ug}
  M.~G.~Alford, S.~Chandrasekharan, J.~Cox and U.~J.~Wiese,
  Nucl.\ Phys.\ B {\bf 602}, 61 (2001)
  [arXiv:hep-lat/0101012].

\bibitem{Kim:2005ck}
  S.~Kim, Ph.~de Forcrand, S.~Kratochvila and T.~Takaishi,
  PoS {\bf LAT2005}, 166 (2006)
  [arXiv:hep-lat/0510069].

\end{thebibliography}
\end{document}